**Nonlinear microwave response of epitaxial YBaCuO films of varying oxygen content on MgO substrates**

M.A. Hein[1], R.G. Humphreys[2] and P.J. Hirst[2], S.H. Park[3] and D.E. Oates[3].

[1] Dept. of Physics, University of Wuppertal, Germany.

[2] QinetiQ Malvern, Worcestershire, UK.

[3] MIT Lincoln Laboratory, Lexington, MA, USA.

Suggested running head: Nonlinear microwave response of YBaCuO on MgO

Abstract − We have investigated the nonlinear microwave properties of electron-beam coevaporated $YBa_2Cu_3O_{7-x}$ films on MgO, using stripline resonators at 2.3 GHz and temperatures 1.7-80 K. The oxygen content of the films ranged from strongly underdoped to overdoped. Above 20 K, the nonlinear response of the resonators was dominated by the superconductor. We could establish clear correlations between the nonlinear surface resistance, two-tone intermodulation (IMD), and oxygen content of the films, which indicate that the superconducting order parameter is important for the nonlinearities. An exponential rather than a power-law representation of the nonlinear current-voltage relation would be required to explain our results phenomenologically. Below 20 K, the dielectric loss tangent of MgO dominated the nonlinear response of the resonators. With increasing power, the dissipation losses decreased markedly, accompanied by enhanced IMD. The surface reactance passed through a shallow minimum at about 5 K, independent of power. We attribute these effects to resonant absorption by impurity states in MgO.

Author to whom communications should be sent:

Dr. Matthias Hein

Dept. of Physics

University of Wuppertal

Gauss-Strasse 20

D-42097 Wuppertal, Germany.

Telephone: +49 (202) 439-2747

Telefax: +49 (202) 439-2811

Email: mhein @venus.physik.uni-wuppertal.de

1. INTRODUCTION

While passive high-temperature superconductor (HTS) microwave devices are finding applications in systems for wireless communication, signal recognition, and medical diagnostics [1], the nonlinear surface impedance $Z_s(I_{rf})$ and third-order intermodulation distortion (IMD) limit wider application. This limitation applies especially to resonant devices in which the oscillating power $P_{osc}$ is high, due to, e.g., enhanced group delay, high quality factors, or high operating power [2]. Understanding and controlling the nonlinear microwave response of HTS films is, therefore, important for narrowband, sharp cut-off, receive filters, which are susceptible to intermodulation, and high-power transmit filters, which suffer from power-dependent absorption. The sources of nonlinear $Z_s$ and IMD, while related, are not well understood at present. In addition, the possible contribution of the substrates to the nonlinear response has been widely neglected until recently [3], although the dielectric loss tangent and permittivity are important parameters for the design and operation of compact microwave devices [2,4].

We have investigated the nonlinear surface impedance $Z_s(I_{rf})$ and IMD of epitaxial $YBa_2Cu_3O_{7-x}$ (YBCO) films of different oxygen deficiency $x$, using stripline resonators at ~2.3 GHz and temperatures 1.7-80 K. We have established clear correlations between the onset of nonlinear $R_s$, the dependence of the IMD power $P_{IMD}$ on $P_{osc}$, and the oxygen content of the YBCO films. The results help to understand the nonlinear response theoretically, by pointing to quasiparticle scattering from defects and pair breaking. In addition, we have identified a nonlinear loss mechanism in MgO, which becomes prominent below about 20 K. This result may have important implications for the fabrication and use of MgO, which has become the substrate material of choice for HTS microwave applications.

2. EXPERIMENTAL TECHNIQUE

2.1 Film Preparation

The 350-nm-thick single-sided epitaxial YBCO films were electron-beam coevaporated at 690°C, close to the 1:2:3 cation stoichiometry, onto commercially available 1cm×1cm×0.5mm MgO substrates [5,6]. The oxygenation level of the as-grown films was adjusted by sequential annealing in argon or plasma-activated oxygen, and characterised by DC magnetisation measurements of the critical temperature $T_c$ and current density $J_c$. The doping level could be inferred from the qualitatively different dependences of $T_c$ and $J_c$ on the oxygen deficiency $x$. In addition, the $c$-axis lattice parameter was determined by the usual $\theta$-$2\theta$ X-ray scans. While the $c$-axis parameter depends on the substrate and film deposition technique, it has been found to be a reliable indicator of relative changes of the oxygenation level under otherwise unchanged deposition conditions [7]. The data obtained for nine YBCO films from three batches are summarised in table I. The films of each batch were deposited simultaneously in the same run, ensuring equivalent deposition and growth conditions. The final oxygenation level of the films ranged from strongly underdoped (strongly reduced $T_c$, low $J_c$) to overdoped (moderately reduced $T_c$, high $J_c$).

To separate dissipation losses in the superconductor from losses in the dielectric substrate, 400-nm-thick Nb films were fabricated for reference measurements [3]. These films were magnetron sputtered at ambient temperature onto 2-inch-diameter MgO wafers [8] and subsequently diced into 1cm×1cm pieces.

2.2 Microwave Measurements

After patterning the superconductor films into 150-μm-wide meander lines by wet chemical etching, they were clamped with two equivalent ground planes to form a stripline resonator [9]. The quality factor $Q$ and resonant frequency were measured at the fundamental resonance $f = \omega/2\pi$ = 2.27 GHz for input power levels $P_{in}$= −85 to +30 dBm. For the fixed geometry of the input and output coupling capacitors, this power range corresponds to electric fields between the center strip and ground planes of $E_{rf} = 0.5 - 5\times10^5$ V/m, or to total microwave currents flowing along the superconducting strip of $I_{rf} = 3\times10^{-6} - 3$ A.

The effective surface resistance derived from the unloaded quality factor comprises losses in the superconductor and the dielectric and is given by [2]:

$$R_{eff} = R_s + G \times \tan\delta, \qquad (1)$$

where $R_s$ is the surface resistance of the superconductor, $\tan\delta$ the loss tangent of the dielectric, and $G$ a geometry factor, which depends on the ratio of penetration depth to film thickness [9,10] and, hence, weakly on temperature ($G \sim 0.8$ Ω is typical for $T < T_c/2$). Changes of the effective reactance are similarly composed of changes of the penetration depth and dielectric permittivity [2]:

$$\Delta X_{eff} = \Delta X_s + G \times \frac{\Delta\varepsilon}{\varepsilon}, \qquad (2)$$

where $\Delta X_s = \mu_0\omega\Delta\lambda$, $\lambda$ is the temperature-dependent superconducting penetration depth and $\varepsilon$ the dielectric permittivity of the substrate.

IMD measurements with a noise floor of $-130$ dBm were performed with two tones of equal input power, separated by 10 kHz. This frequency spacing approached the full width at half maximum of the resonance for the highest measured Q-values, but it is considered still narrow enough to provide a reliable qualitative picture of the IMD behaviour. More accurate modeling of the nonlinear response including the bandpass characteristic of the resonator is in progress.

## 3. NONLINEAR MICROWAVE RESPONSE OF YBCO ($T \geq 20$ K)

### 3.1 Power Handling and Oxygenation Level

Figure 1 compares the dependence of the surface resistance on the total RF current for three YBCO films with different oxygen deficiencies, all from batch #3, at two representative temperatures $T = 20$ K and 50 K. The plots show how the onset for a current-dependent $R_s$ moves to lower currents, and the steepness of $R_s(I_{rf})$ increases, as the films become more oxygen deficient. The steep onset of the nonlinear response is displayed on logarithmic scales in panel *b*. This plot also illustrates the high quality of the films studied. The surface resistance of, e.g., the overdoped film #3D at 20 K remains constant at $R_s \sim 4$ µΩ up to $I_{rf} \sim 1$ A, which corresponds to a magnetic flux density of about 30 mT. This value is only about a factor of 2.5 below the lower critical field expected for bulk YBCO with the RF currents flowing entirely within the Cu-O planes [2].

The RF current dependence of the surface reactance, $\Delta X_s(I_{rf})$, displayed qualitatively the same behaviour as the surface resistance, with the differential loss tangent $r \equiv (\partial R_s/\partial I_{rf})/(\partial X_s/\partial I_{rf})$ approaching at high power constant values between about 1 and 3. The constancy of $r$ indicates that both $R_s$ and $X_s$ follow the same functional dependence on $I_{rf}$ at sufficiently high power.

The observed correlation between oxygen deficiency and nonlinear surface impedance agrees with the common expectation that fully oxygenated films display superior electronic performance, and has accordingly been speculated for some time. However, due to the complicated interdependence of deposition technique, film morphology, and oxygenation level, it has neither been studied systematically nor observed until now.

3.2 Scaling of the Power Handling, Critical RF Current

In order to explore the nature of the nonlinear mechanism in greater detail, it is necessary to distinguish between qualitative changes of the $R_s(I_{rf})$-dependence brought about by different oxygenation levels, and mere changes of the current scales, on which the nonlinearities occur. For this purpose, Fig. 2 (a) shows typical $R_s(I_{rf})$-data for a set of temperatures. Figure 2 (b) shows the same data after re-scaling $I_{rf}$ at every temperature such that the scaled $R_s(I_{rf}^*)$-curves merged onto, say, the data for $T = 20$ K. The obvious success of this procedure demonstrates that the temperature dependence of the nonlinear $Z_s$ is caused by that of the scaling current alone. The data in Fig. 2 (b) are consistent with a quadratic current dependence, e.g., in form of:

$$R_s(I_{rf}, T) = R_{s0}(T) \text{ for } I_{rf} < I_0, \text{ and}$$

$$R_s(I_{rf}, T) = R_{s0}(T) + \alpha \times \left[\frac{I_{rf}}{I_0(T)} - 1\right]^2 \text{ for } I_{rf} \geq I_0, \quad (3)$$

where $\alpha$ is a constant and $I_0$ defines the appropriate current scale. The two parts of Eq. (3) emphasise the existence of a threshold current, below which $R_s(I_{rf})$ is constant. We note that at high currents $I_{rf} > I_0$ Eq. (3) is consistent with the simple quadratic approach $R_s = R_0 \times (1 + b_R \times I_{rf}^2)$ used in previous work [11]. The current-scaling of the $R_s(I_{rf})$-dependences of all other samples listed in table I worked equally well, so that we can exclude functional changes of the $Z_s(I_{rf})$-dependence for varying oxygen deficiency $x$.

So far, the discussion has focussed on changes of the current scale relative to that at 20 K. To derive absolute values for the scaling current, one has to define an appropriate criterion. We have found the following, implicit, definition useful for both low and high-$T_c$ superconductors [12]:

$$R_s(I_{crit}) \equiv 2.5 \times R_{s0}(I_{rf} \to 0) . \tag{4}$$

Using $I_{crit}$ as a criterion for a critical RF current is more versatile than using $I_0$ in Eq. (3). Equation (4) includes a possible variation of the surface resistance at low currents, $I_{rf} < I_0$, and combines the two parameters $\alpha$ and $I_0$ into the single parameter $I_{crit}$. There are obviously more possibilities to define a characteristic current scale, which, however, will not alter the conclusions to be drawn.

Figure 3 displays the results for $I_{crit}$ versus reduced temperature $t \equiv T/T_c$, for all films studied. The properties of the samples discussed so far provide insufficient information to relate the absolute $I_{crit}(0)$-values to the film preparation. However, almost all curves have in common a nearly linear temperature dependence of $I_{crit}$ for $T/T_c > 0.5$. Such a behaviour is expected and was observed for various types of grain-boundary Josephson junctions, where the coupling between the superconducting electrodes is believed to occur via localized states [13,14]. However, in view of the high quality of the films (high $J_c$-values and good power handling), we don't expect granular effects to be relevant in our samples. Another, more likely, explanation is based on the observation of a linear temperature dependence of the magnetic field at which flux penetrated a superconducting stripline [15]. We are thus led to speculate that the RF critical current $I_{crit}$ could be related to the DC critical current density $J_c$. This is indeed the case, as illustrated in more detail in the following section.

3.3 Correlations between DC and RF nonlinearities

The critical current density $J_c$ is a key parameter to describe the nonlinear current-voltage characteristics (IVC) of superconductors at DC or low frequencies. We have therefore examined our microwave data for correlations between characteristic RF parameters and $J_c$. The first parameter considered is the critical RF current introduced by Eq. (4). The other two parameters are derived from our results for IMD as follows:

1. The third-order intercept (TOI) quantifies the power level at which the extrapolated IMD signal equals the input power. To avoid complications related to the geometry of our resonators and the nonlinear compression of $P_{out}(P_{in})$, we have derived TOI-data from plotting the power measured at the two symmetric third-order IMD-sidebands, $P_{IMD}$, versus output power at the fundamental tone, $P_{out}$. We note that $P_{out}$ is proportional to the oscillating power. Under certain experimental conditions, the measured $P_{IMD}(P_{out})$-curves were not simple straight lines but displayed curvature (for an example, see Fig. 6 and related discussion in Sec. 4). The TOI was determined in such cases from the asymptotic behaviour of $P_{IMD}$ at high power levels (e.g., above –20 dBm in Fig. 6). TOI-values could be derived graphically or numerically, according to the condition $P_{IMD} = P_{out}$, using the entire set of IMD data, or just one point on the $P_{IMD}(P_{out})$-curve and the slope in that point. Both techniques yielded almost identical results (filled and open symbols in Fig. 4b), confirming the consistency of our analysis. If the IV-relation effective at microwave frequencies were described by a power series expansion, the TOI (in dBm) should scale logarithmically with the square of the characteristic current [2,16].

2. The slope $m$ of the $P_{IMD}(P_{out})$-curves, plotted on logarithmic scales, measures the exponent of the functional dependence of $P_{IMD}$ on $P_{out}$, $m = \partial \log P_{IMD} / \partial \log P_{out}$. In case of a power-series representation of the IVC, third-order effects like frequency intermodulation or harmonic generation lead to a constant slope, $m = 3$.

Figure 4 summarises our results for $I_{crit}$(50K), TOI(20K), and $m$(40K) versus the critical DC current density $J_c$(60K). The different temperatures at which these parameters were evaluated are due to experimental considerations. They do not affect our conclusions, especially since we don't attempt a quantitative analysis of our results. The diagrams of Fig. 4 have two features in common: 1. We observe clear correlations between all three RF parameters and $J_c$, indicating that the physical significance of $J_c$ persists well into the microwave region. 2. The samples of batch #1 fall apart from the general trends. This important result proves that good RF performance and good DC performance are not equivalent. We attribute the out-of-sequence results for batch #1 to the different annealing parameters compared to those used for batches #2 and #3. The annealing temperatures for the films of batch #1 remained below 400 $^o$C, and the final oxygenation level was approached from the as-grown state. In contrast, the films of batches #2 and #3 were sequentially de- and re-oxygenated by annealing in argon and activated oxygen, at temperatures up to 540 $^o$C. These differences certainly affect the density and homogeneity of oxygen vacancies, and hence the local variation of the superconducting order parameter. We are presently trying to reproduce the results for batch #1 with a new set of films, to achieve deeper insight into the mechanisms which are necessary to achieve good RF performance.

While the $I_{crit}$-$J_c$-correlation (top panel of Fig. 4) conforms to our intuitive expectation, the TOI-$J_c$-correlation (middle panel) is quite surprising. The TOI, expressed in units of dBm, should depend on the *logarithm* of $J_c$, if the IV-relation were given by a power-series. This is in contrast with the *linear* dependence of the TOI on $J_c$ we observe. Our result implies that a simple power-series expansion with real coefficients is not sufficient to describe the nonlinear microwave response of HTS films. A possible solution of this dilemma is the assumption of an exponential IV-dependence like $V_{rf}(I_{rf}) \propto \exp(I_{rf}/I_{scale})$, which could be associated with flux creep. However, such a speculation lacks theoretical justification at present. More sophisticated models of the IMD behaviour are also under development.

Another surprising result is the systematic variation of the slope $m$ with $J_c$ (bottom panel of Fig. 4), which indicates that the IMD slope decreases for increasing sample quality. Such a variation of $m$ is, again, not compatible with a IV-relation given by a power-series. Furthermore, slopes $m > 3$ seem to be possible (#3A and #3C), while the most strongly overdoped film #2C shows $m \sim 2$. A quadratic power dependence of $P_{IMD}$ was previously observed for TlBaCaCuO films [17]. Such a behaviour can result, theoretically, from a nonlinear surface impedance that depends on the modulus of the RF current, in sharp contrast to the quadratic behaviour discussed in Sec. 3.2 [17,18]. The only physical mechanism known at present, which could lead to $\Delta Z_s(I_{rf}) \propto |I_{rf}|$, is the Yip-Sauls regime of d-wave superconductors, which can be experimentally accessed only at very low temperatures [18]. The persistence of $m < 3$ up to such high temperatures like 40 K is presently not understood but definitely deserves further attention.

## 4. NONLINEAR MICROWAVE RESPONSE OF MgO ($T < 20$ K)

In Sec. 3, $T = 20$ K was the lowest temperature analysed. The reason for this choice is the occurrence of unexpected microwave properties below 20 K, which are associated with the nonlinear dielectric microwave impedance of MgO [3,19],

$$Z_d = \sqrt{\frac{\mu_0}{\varepsilon_0 \varepsilon_r}} \times \left(1 + \tfrac{1}{2} i \tan \delta\right). \tag{5}$$

As a consequence, the effective surface impedance $Z_{eff}$, according to Eqs. (1) and (2), can no longer be identified with $Z_s$ of the superconductor, as in the previous section. Since our first observation and tentative interpretation in [3], we have refined our analysis and complemented by new results, as described in the following three subsections.

### 4.1 Temperature Dependence

Figure 5 summarises typical results for the temperature dependence of $Z_{eff}$, measured for YBCO films on MgO. The different symbols represent the asymptotic behaviour observed at very low and moderate power levels, respectively, as illustrated in more detail by the circles in Fig. 6. $R_{eff}(T)$ decreased at the higher power and approached 2 µΩ. Expecting $R_s \geq 1$µΩ for high-quality epitaxial YBCO films [2], we estimate from the $R_{eff}$(20K)-value an upper-limit for the loss tangent of about $10^{-6}$, consistent with measured tan$\delta$- data for MgO at 20 K [20,21]. At very low power, $R_{eff}(T)$ increased and became dominated by dielectric losses. This could be confirmed by the very similar temperature and power dependences of $R_{eff}$ for Nb on MgO (see [3] and Figs. 7-9 below). Such an agreement for the two totally different superconductors on MgO, and the absence of this nonlinear behaviour for other substrates like sapphire and LaAlO$_3$, proves that our observations are caused by the dielectric response of MgO. Using $R_s \sim 2$ µΩ for YBCO and the effective value, $R_{eff} \sim 20$ µΩ, we derive tan$\delta$(1.7K) $\approx 2\times 10^{-5}$, which is about 20 times above tan$\delta$(20K) (c.f., Fig. 7 (a)).

Complementary to the loss tangent, the dielectric permittivity, monitored by changes of the effective surface reactance $\Delta X_{eff}(T)$, displayed a shallow minimum at $T \sim 5$ K, but remained independent of power up to moderate power levels (Fig. 5 (b) and squares in Fig. 6). Reflecting again the growing influence of $Z_d(T)$ on $Z_{eff}(T)$ at low temperatures, the increase of $\Delta X_{eff}(T)$ in Fig. 5 can be entirely associated with the dielectric permittivity of MgO. Such an extremal behaviour must not be confused with anomalous effects in the cuprate superconductor, which could be misinterpreted, e.g., as an indication of Andreev bound states at interfaces in d-wave superconductors [22].

4.2 Power Dependence

Figure 6 compares typical results for the power dependence of $R_{eff}$, $\Delta X_{eff}$, and $P_{IMD}$ at $T = 5$ K, here plotted versus input power. The $R_{eff}$ decreased markedly as $P_{in}$ increased, for the given resonator geometry and coupling above −60 dBm, reached a plateau around −40 dBm, and increased again at around −10 dBm. The difference between the low-power value of $R_{eff}$ and its minimum increased strongly for decreasing temperature, in accordance with Fig. 5 (a). In contrast to the resistance, $\Delta X_{eff}$ was independent of power at low power and increased smoothly above −30 dBm. The increase of $R_{eff}$ and $\Delta X_{eff}$ in the power range above −10 dBm is attributed to the nonlinear response of YBCO (Sec. 3). The IMD signal passed through a plateau in the region where $R_{eff}$ decreased, and rose again more steeply where both $X_{eff}$ and $R_{eff}$ increased. It is worth noting that the frequency intermodulation, caused by the nonlinear loss tangent of MgO, is intimately related to $R_{eff}$. This is qualitatively different from the IMD behaviour expected for the superconductor at high power or at $T > 20$ K, where $P_{IMD}$ is usually dominated by the nonlinear reactance. Obviously, the dielectric microwave impedance of MgO induces enhanced frequency intermodulation, which could strongly degrade the performance of microwave devices.

4.3 Theoretical Description and Comparison between YBCO/MgO and Nb/MgO

The unique temperature and power dependences of the dissipative and reactive response of MgO (Figs. 5 and 6) helped us to identify the underlying mechanism. Very similar features were observed for amorphous glasses and highly disordered crystalline solids in similar ranges of frequency and temperature [23,24]. The observed behaviour was attributed to tunneling of atoms or molecules between nearly equivalent positions in the solid. This scientific area is subject to ongoing experimental and theoretical research [25]. Nevertheless, many features can be understood in the framework of a simple quantum mechanical two-level system. Phenomenologically, this model has similarities with the double-well potential underlying the Debye equations for dielectric relaxation [3,26]. The major difference between the two models is the resonant character of the transitions in the former, in contrast to the relaxational character in the latter. Dielectric resonances of isolated defects usually occur in the sub-mm wave range or at even higher frequencies, but recent theoretical studies have shown that clustering of defects can reduce the resonant frequencies very strongly, so that they may even reach into the microwave range [27].

The temperature dependence of the loss tangent can be easily derived from Fermi's golden rule. The temperature dependence of the dielectric permittivity, then, follows from that of tan$\delta$ through the Kramers-Kronig-relation, which requires additional information on the frequency range over which resonances occur. Different results have to be expected for solids with rather dilute defects, where the resonances remain sharp and limited to a well-defined frequency interval, and for highly disordered systems. According to ref. [24], one gets

$$\tan \delta(T) = \frac{\pi}{\varepsilon_0 \varepsilon_r} \times n_d \mu^2 \times \tanh\left(\frac{hf}{2kT}\right) \quad (6)$$

for the dielectric loss tangent at low intensities, and

$$\frac{\Delta \varepsilon_r(T)}{\varepsilon_r} = -\frac{2}{\varepsilon_0 \varepsilon_r} \times n_d \mu^2 \times \ln\left(\frac{T}{T_0}\right) \quad (7)$$

for the fractional change of the permittivity in highly disordered systems. In Eqs. (6) and (7), $n_d$ and $\mu$ describe the density and the electric dipole moment of the defects.

The two-level model also explains in a natural way the anomalous power dependence, as resulting from saturation effects due to enhanced population of the higher energy level for increasing intensity of the microwave field. To derive the power dependence of the complex permittivity requires a more sophisticated mathematical framework, which bears similarities with the Bloch-equations for a spin-1/2-particle in a magnetic field [23]:

$$\tan \delta(S_{rf}) = \tan \delta(0) \times \left(1 + \frac{S_{rf}}{S_0}\right)^{-1/2}, \quad (8)$$

where $S_{rf}$ denotes the intensity of the RF field, $S_{rf} \propto E_{rf}^2$, and the real part of the permittivity is independent of power. The threshold intensity $S_0$ depends on $\mu$ but not on $n_d$. It bears also information on the characteristic relaxation times of the excited defect states [24]. In ref. [3], the interpretation of the anomalous power dependence of $\tan \delta$ in terms of a microwave field-dependent dielectric relaxation time was obviously oversimplified. Equation (8) provides a more accurate explanation, connecting the experimental data now with a microscopic mechanism.

We compare our experimental results with the theoretical expectation in Figs. 7 and 8. In Fig. 7 (a) we have plotted the difference between the two curves shown in Fig. 5 (a), normalised by G, versus the temperature function expected from Eq. (6). We have also included in this figure our results for a Nb film on MgO (diamonds). Both data sets are consistent with the expected linear relationship. The similar slopes indicate very similar properties, $n_d\mu^2$, of the MgO substrates, although these were bought from different suppliers. Fig. 7 (b) shows the corresponding result for the fractional variation of $\varepsilon_r$ expected from Eq. (7). The data derived for the YBCO film were not corrected for changes of the superconducting penetration depth, since $\lambda(T)$ can be safely assumed to be constant below 20 K. The situation is different for Nb with $T_c = 9.2$ K, where we accounted for $\Delta\lambda(T)$ according to BCS theory. As for the loss tangent, we find very similar results for both superconductors. The slopes in Figs. 7 (a) and 7 (b), which should differ for a highly disordered system by a factor of $-\pi/2 \sim -1.6$, differ by $-5.5\times10^{-4} / 6.5\times10^{-5} \sim -8.5$. The difference of a factor of $\sim 5$ is attributed to the nature of the defects in a highly crystalline material like MgO.

A typical measured power dependence of $\tan\delta$ is compared with Eq. (8) in Fig. 8 for a YBCO film and a Nb film, both on MgO. The symbols represent the measured data, the curves denote the result of a single-parameter fit

$$R_{eff}(E_{rf}) = R_s + \Delta R_{eff} \times \left(1 + \frac{E_{rf}^2}{E_0^2}\right)^{-1/2}, \qquad (9)$$

which is equivalent to Eq. (8). The parameters $R_s$ and $\Delta R_{eff}$ are given by the experimental data, while $E_0$ can be varied within a narrow range to optimise the fit. The data in Fig. 8 indicate similar $E_0$-values for the two types of films on MgO substrates used, which implies similar types of defects. The $\Delta R_{eff}$-values for Nb/MgO seem a bit higher than for YBCO/MgO, which could hint for a larger defect density in the substrate.

Finally, in Fig. 9, we summarise the temperature dependences of $\Delta R_{eff}$ (panel *a*) and $E_0$ (panel *b*) for all samples studied. Figure 9 (a) illustrates the increase of the loss tangent for decreasing temperatures. All samples lead to comparably high and temperature dependent $\Delta R_{eff}$-values, except for two films of batch #1 (C and D), and the as-grown film #3A. It is tempting to attribute this behaviour to the different annealing schedules (see discussion in Sec. 3.3), as these can well change the density of defects in MgO. Possible candidate defects were identified in [19] to be OH$^-$–ions, which can be released from grain boundaries upon heating. Further studies are planned to verify our assumption.

The temperature variation of $E_0$ displayed in Fig. 9 (b) results mainly from the temperature dependent relaxation times of the defects in MgO. There is only one exceptional result among all others, namely for #1B. The reason for this peculiarity is not known at present. Complementary experiments like specific-heat measurements are required to extract more detailed information on the defect density and, in turn [24], the relaxation times of the defects in MgO that cause the nonlinearities.

## 5. CONCLUSIONS

We have investigated the nonlinear microwave properties of electron-beam coevaporated YBa$_2$Cu$_3$O$_{7-x}$ films on MgO of varying oxygen content, at 2.3 GHz and temperatures 1.7-80 K. Above 20 K, the nonlinear response of the resonators was dominated by the HTS films for both substrate materials. We have established clear correlations between the onset of nonlinear $R_s$, the dependence of $P_{IMD}$ on the oscillating power $P_{osc}$, and the oxygen deficiency $x$ of the YBCO films. With increasing $x$, the critical currents and third-order intercepts decreased, while the exponent of the power-law relation between $P_{IMD}$ and $P_{osc}$ increased from 2 to 3. This observation defines a route to optimise the linear response of HTS films for microwave applications.

In MgO, at temperatures below 20 K, the dielectric loss tangent was found to dominate the nonlinear response of the resonators. The dissipation losses decreased with increasing power by up to one order of magnitude. This anomaly was accompanied by enhanced IMD. The surface reactance passed through a shallow minimum at about 5 K, but remained independent of power. The results were very similar for all investigated YBaCuO films as well as for Nb films on MgO made for reference. We attribute these effects to resonant absorption by impurity states in MgO. While proper identification of the impurities is lacking at present, $OH^-$–ions are considered the most likely candidates, since they are greatly abundant in melt-fused MgO.

ACKNOWLEDGEMENTS

We gratefully acknowledge valuable contributions from J. Derov, P. Droste, T. Dahm, C. Enss, B. Henderson, S. Hunklinger, K. Peters, and D. Seron. The work at QinetiQ was supported by the MOD (UK). The work at Lincoln Laboratory was supported by the AFOSR (U.S.A.). The work at Wuppertal was supported by the European Office of Aerospace Research and Development, AFOSR/AFRL, under Contract No. F61775-01-WE033.

TABLES

Table I. Overview of the YBaCuO films on MgO: $T_c$ = critical temperature, $J_c$ = critical current density at 60 K, $\Delta c$ = deviation of the c-axis lattice parameter from the value 1.16 nm. All but the as-grown films were post annealed, in successive steps, in argon or plasma-activated oxygen. The oxygenation level of the as-grown films was near optimum doping.

| Sample Batch / ID | $T_c$ (K) | $J_c$(60K) (MA/cm$^2$) | $\Delta c$ (pm) | Oxygen doping level |
|---|---|---|---|---|
| 1B | 90.45 | 2.67 | 11.5 | as-grown |
| 1C | 90.20 | 2.23 | 11.6 | under |
| 1D | 89.00 | --- | 10.3 | ~ optimum |
| 2A | 91.74 | 11.40 | 10.3 | ~ optimum |
| 2B | 90.68 | 5.82 | 12.5 | under |
| 2C | 84.18 | 10.40 | 8.6 | over |
| 3A | 90.54 | 2.40 | 13.4 | as-grown |
| 3C | 69.3 | 2.67 | 15.2 | under |
| 3D | 91.80 | 8.10 | 10.2 | over |

FIGURES AND FIGURE CAPTIONS

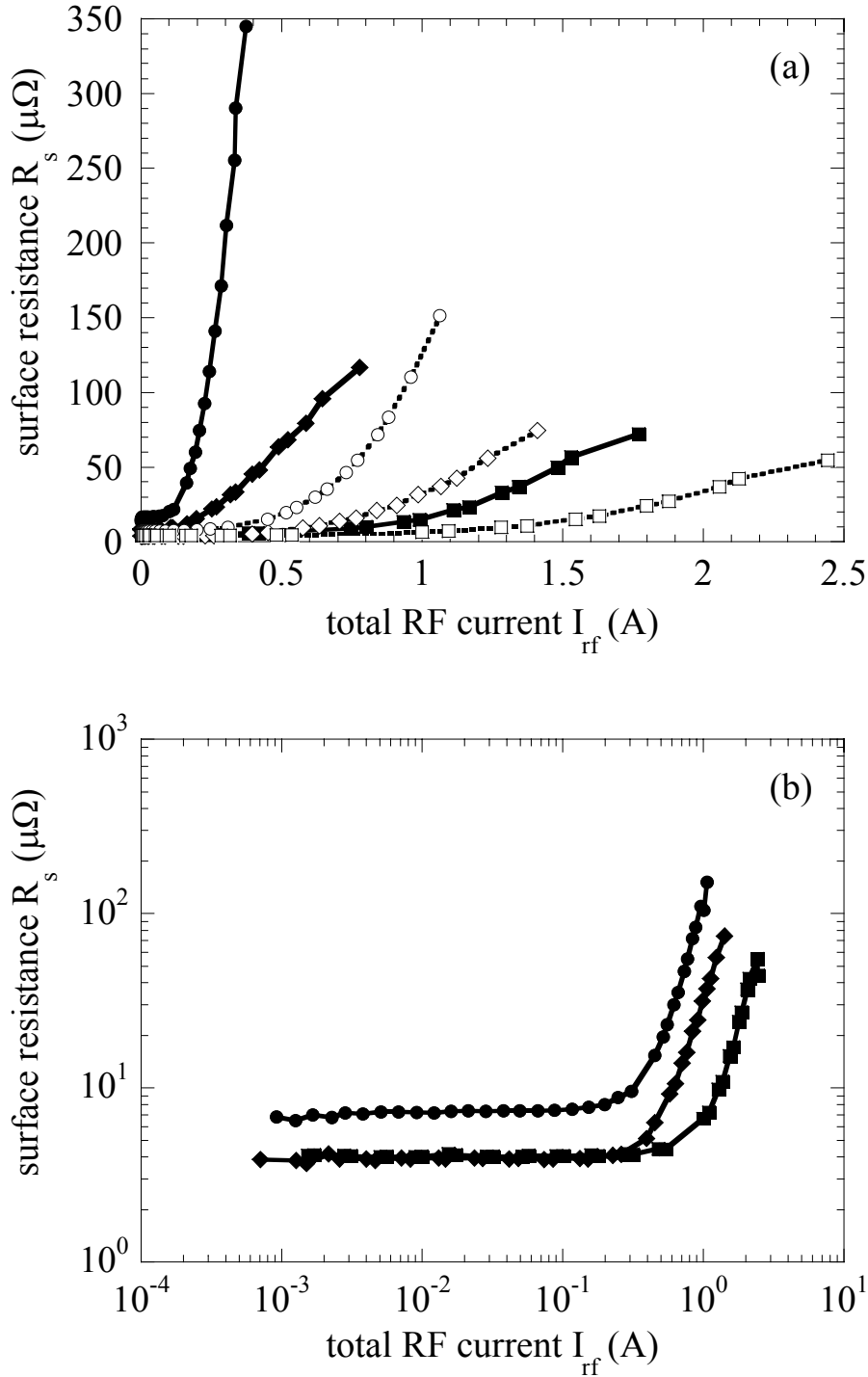

Fig. 1. RF-current dependence of $R_s$ for the samples #3A (near optimum doping, diamonds), #3C (underdoped, circles), and #3D (overdoped, squares) at $f$ = 2.27 GHz and $T$ = 20 K (open symbols, dotted interpolation curves) and 50 K (full symbols, solid curves), using linear scales (panel *a*) and logarithmic scales (panel *b*, $T$ = 20 K only).

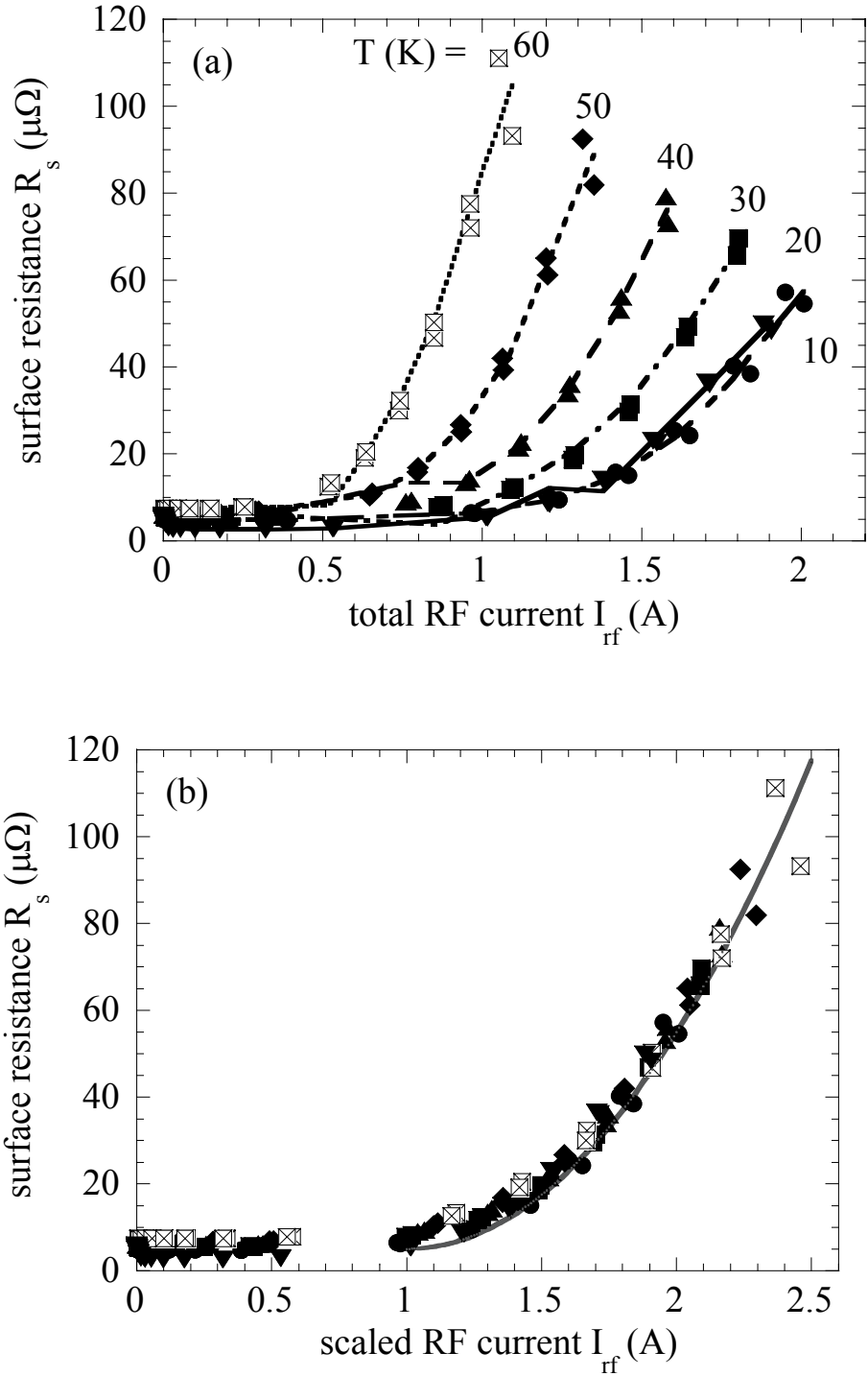

Fig. 2. (*a*) RF-current dependence of $R_s$ for sample #2B at $f = 2.27$ GHz and various temperatures as indicated (different symbols and interpolation curves). (*b*) Same data after scaling $I_{rf}$ such that all curves merge with the results for $T = 20$ K. The solid line represents the quadratic fit of Eq. (3).

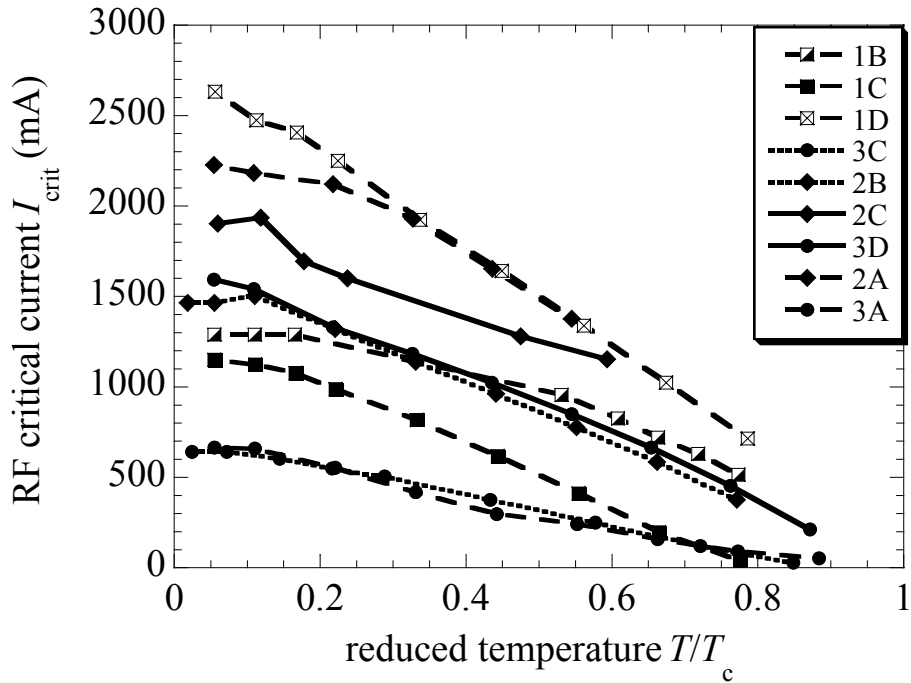

Fig. 3. Critical RF-current versus reduced temperature at $f$ = 2.3 GHz for all investigated YBCO films (for details see legend): batch #1 (squares), #2 (diamonds), #3 (circles); underdoped (dotted interpolation curves), near optimum doping (dashed), overdoped (solid).

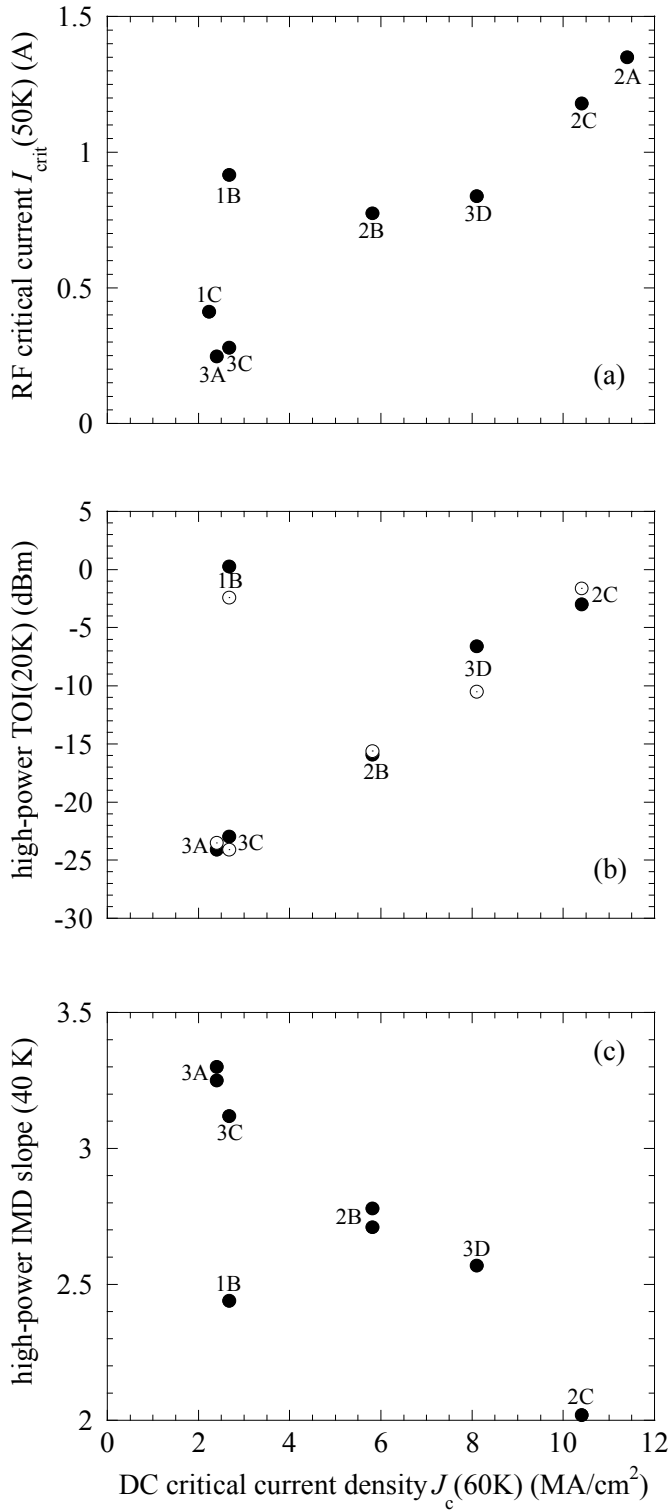

Fig. 4. Correlation between the DC critical current density $J_c$(60K) and the RF critical current $I_{crit}$(50K) (panel *a*), high-power third-order intercept TOI(20K) (panel *b*), and high-power IMD slope (40K) (panel *c*) for the samples indicated. The TOI-data were determined numerically (filled symbols) and graphically (open symbols) as explained in Sec. 3.3.

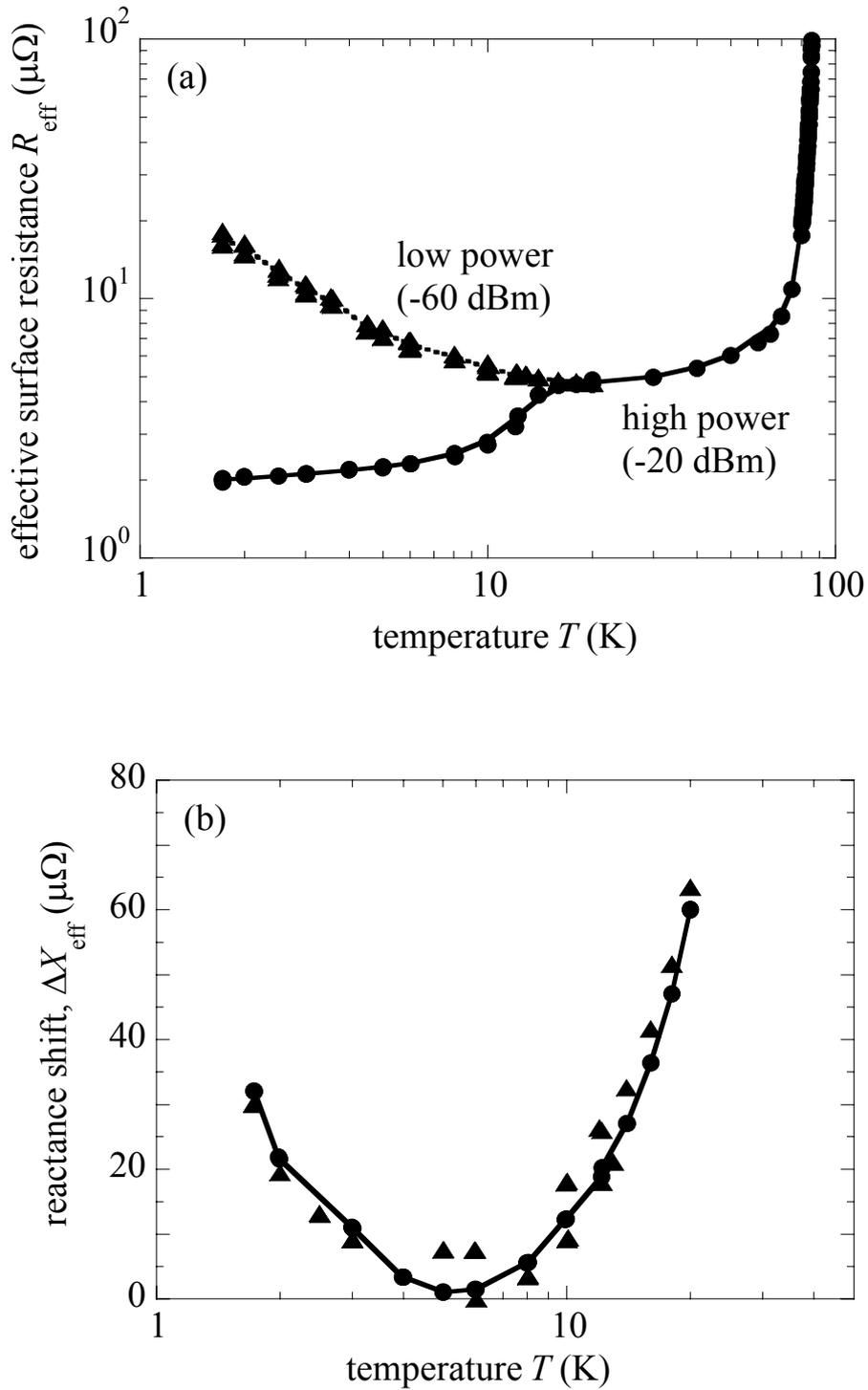

Fig. 5. Typical data for the effective surface resistance (panel *a*) and shift of surface reactance (panel *b*) at 2.3 GHz for YBCO on MgO at temperatures below 20 K. The different symbols and interpolation curves denote extreme power levels, as described in the text.

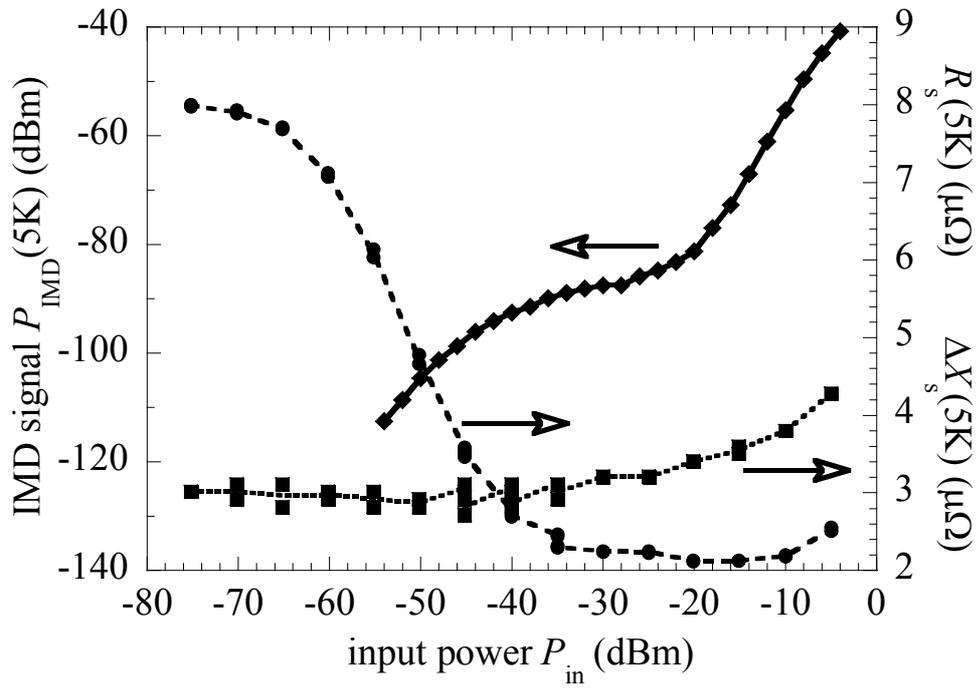

Fig. 6. Typical power dependence of the effective surface resistance (circles, dashed interpolation), change of surface reactance (squares, dotted), and IMD signal (diamonds, solid) for YBCO on MgO (sample #2B) at $f$ = 2.3 GHz and $T$ = 5 K.

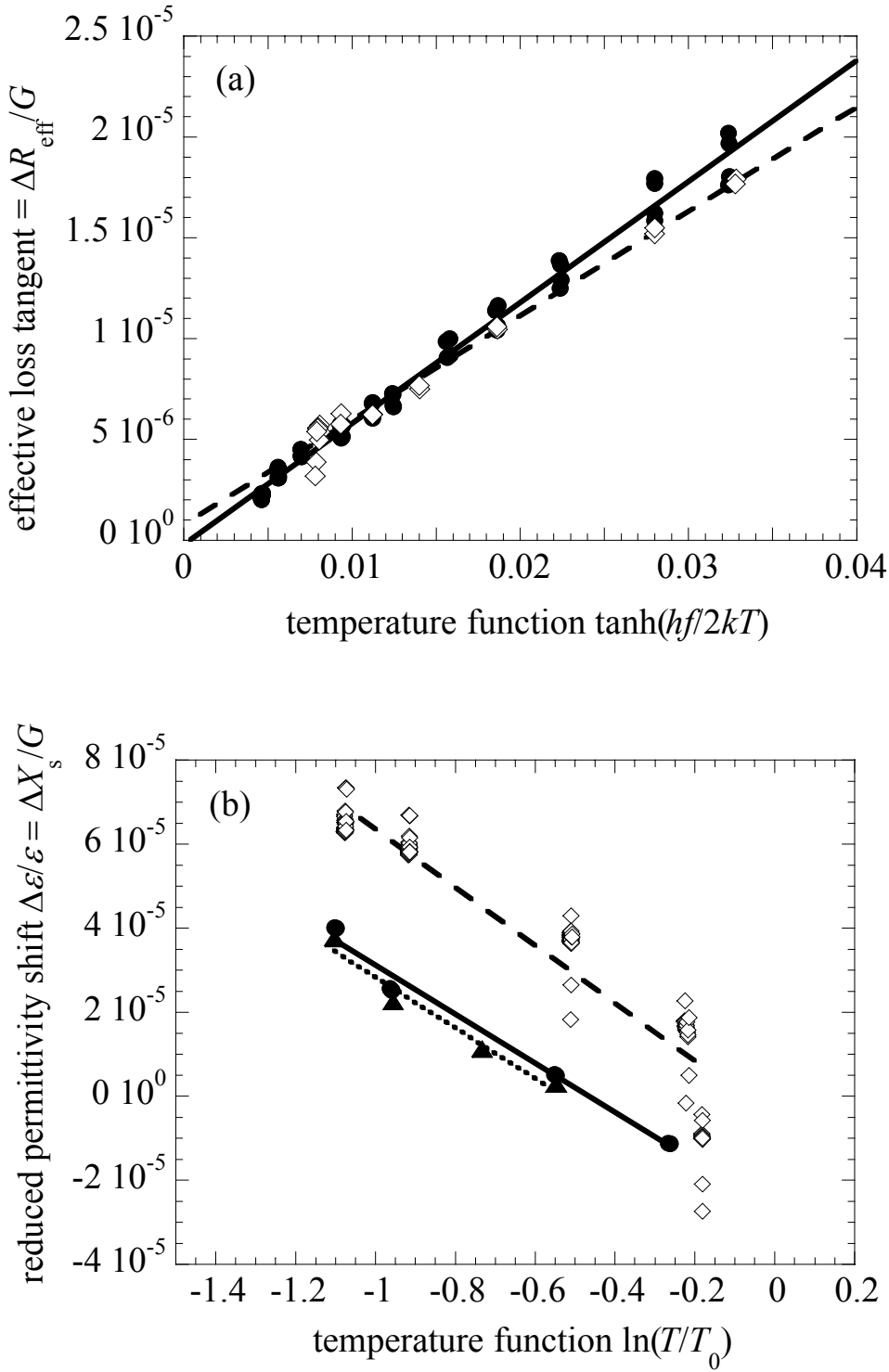

Fig. 7. Temperature dependent dielectric loss tangent (panel *a*) and permittivity (panel *b*) at $f$ = 2.3 GHz for YBCO on MgO ($P_{in}$=-60 dBm: triangles, dotted interpolation, $P_{in}$=-20 dBm: circles, solid) and Nb on MgO (various power levels: diamonds, dashed), as expected for resonant absorption. $T_0$ = 5.2 K. The $\Delta X_{eff}(T)$-data for Nb were corrected for the temperature variation of the superconducting penetration depth according to the BCS theory ($T_c$ = 9.2 K, $\Delta(0)/kT_c$ = 1.8).

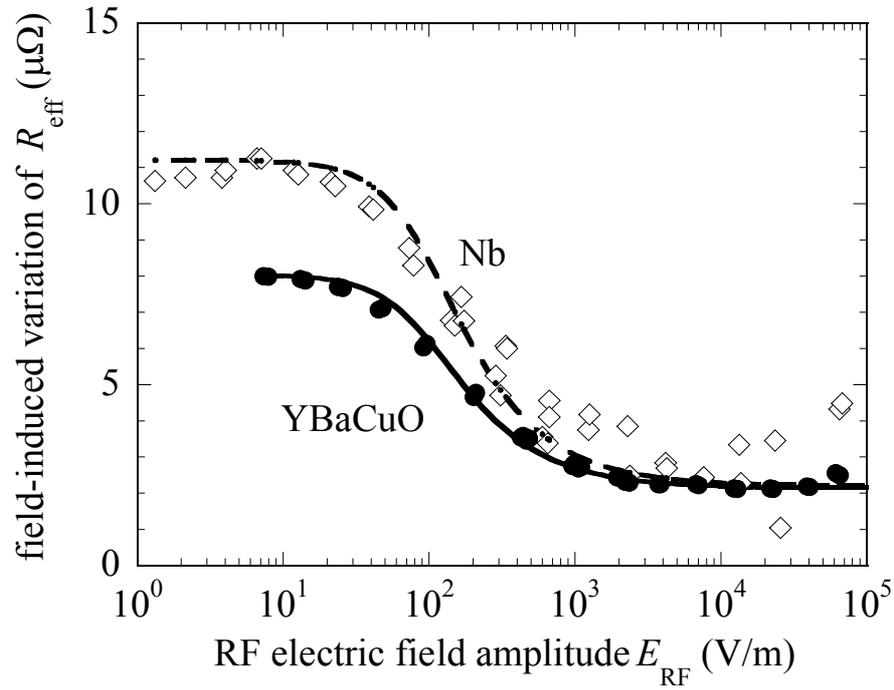

Fig. 8. Electric field-induced variation of the effective surface resistance for YBaCuO (circles) and Nb (diamonds) on MgO at $f = 2.3$ GHz and $T = 5$ K. The dashed and solid curves denote fits to Eq. (9).

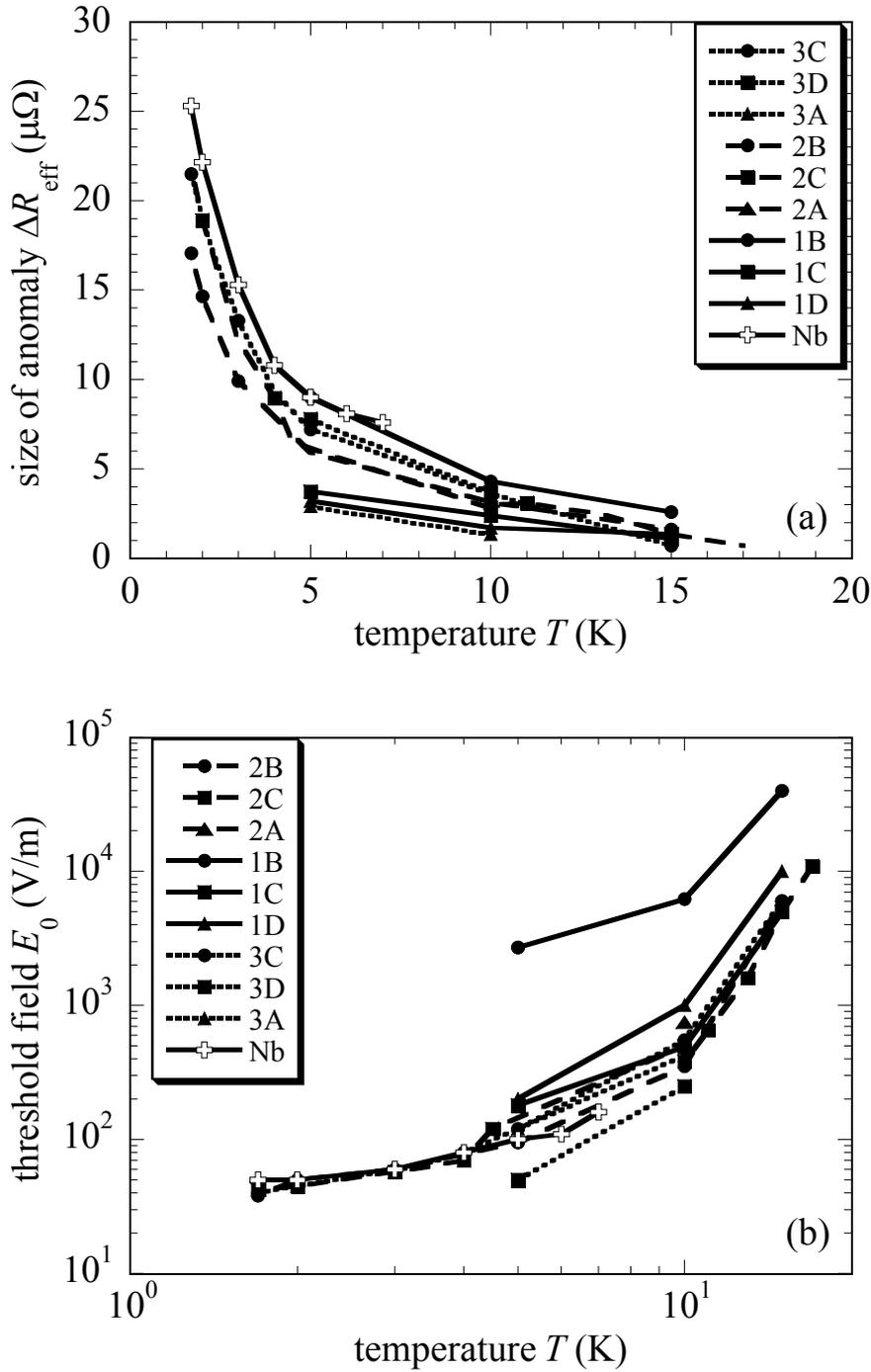

Fig. 9. Temperature dependences of the fit parameters $\Delta R_{eff}$ (panel *a*) and $E_0$ (panel *b*) for the power dependent loss tangent of MgO at $f = 2.3$ GHz and $T = 5$ K for the samples indicated in the legends (symbols and interpolation curves).

REFERENCES

[1] See, e.g., other contributions to this symposium.

[2] M.A. Hein, High-Temperature Superconductor Thin Films at Microwave Frequencies, Springer Tracts of Modern Physics, Vol. 155 (Springer, Heidelberg/New York, 1999).

[3] M.A. Hein et al., Appl. Phys. Lett. 80, 1007 (2002).

[4] M.J. Lancaster, Passive Microwave Device Applications of High-Temperature Superconductors (Cambridge University Press, Cambridge, 1997).

[5] R. G. Humphreys et al., Materials Science and Engineering B10, 293 (1991).

[6] N.G. Chew et al., Appl. Phys. Lett. 57, 2016 (1990).

[7] S. Orbach-Werbig, Dissertation, University of Wuppertal, Report WUB-DIS 94-9, unpublished (1994); N.G. Chew et al., IEEE Trans. Appl. Supercond. 5, 1167 (1995).

[8] The niobium was deposited as described in K.K. Berggren et al., IEEE Trans. Appl. Supercond. 9, 3271 (1999).

[9] D. E. Oates et al., IEEE Trans. Microwave Theory Tech. 39, 1522 (1991).

[10] D. M. Sheen et al., IEEE Trans. Appl. Supercond. 1, 108 (1991).

[11] P.P. Nguyen et al., Phys. Rev. B 48, 6400 (1993).

[12] M.A. Hein, M. Perpeet, and G. Müller, IEEE Trans. Appl. Supercond. 11, 3434 (2001).

[13] A.A. Golubov et al., in Applied Superconductivity 1997, H. Rogalla and D.H.A. Blank, eds, IOP Conference Series, No. 158, (IOP Publishing, Bristol, 1997), pp. 463-466.

[14] J. Halbritter, J. Supercond. 11, 231 (1998).

[15] A.N. Grigorenko et al., Appl. Phys. Lett. 78, 1586 (2001).

[16] J.S. Booth et al., J. Appl. Phys. 86, 1020 (1999).

[17] B.A. Willemsen, K.E. Kihlstrom, and T. Dahm, Appl. Phys. Lett. 74, 753 (1999).

[18] T. Dahm and D.J. Scalapino, Phys. Rev. B60, 13125 (1999).

[19] B. Henderson and W.A. Sibley, J. Chem. Phys. 55, 1276 (1971).

[20] J. Krupka, R. G. Geyer, M. Kuhn, and J. H. Hinken, IEEE Trans. Microwave Theory Tech. **42**, 1886 (1994).

[21] T. Konaka, M. Sato, H. Asano, and S. Kubo, J. Supercond. **4**, 283 (1991).

[22] H. Walter et al., Phys. Rev. Lett. 80, 3598 (1998); A. Carrington et al., Phys. Rev. Lett. 86, 1074 (2001).

[23] S. Hunklinger and M. von Schickfus, in Amorphous Solids, W. A. Phillips, ed. (Springer, Berlin/Heidelberg/New York, 1981), pp. 81-105.

[24] U. Strom, M. von Schickfus, and S. Hunklinger, Phys. Rev. Lett. 41, 910 (1978).

[25] S. Hunklinger and C. Enss, in Insulating and Semiconducting Glasses, P. Boolchand, ed., Series of Directions in Condensed Matter Physics, Vol. 17, p. 499 (World Scientific, Singapore, 2000).

[26] V. V. Daniel, Dielectric Relaxation (Academic Press, New York, 1967).

[27] M.W. Klein, Phys. Rev. B40, 1918 (1989).